# Unexpectedly strong diamagnetism and superparamagnetism of aromatic peptides due to self-assembling and cations

Haijun Yang[1,2], Liuhua Mu[2], Yongshun Song[3], Zixin Wang[4], Xin Zhang[5], Jun Hu[1], Feng Zhang[4,6*], Haiping Fang[3,1*]

1. Shanghai Synchrotron Radiation Facility, Zhangjiang Laboratory (SSRF, ZJLab), Shanghai Advanced Research Institute, Chinese Academy of Sciences, Shanghai 201204, China
2. Division of Interfacial Water, CAS Key Laboratory of Interfacial Physics and Technology, Shanghai Institute of Applied Physics, Chinese Academy of Sciences, Shanghai 201800, China
3. School of Science, East China University of Science and Technology, Shanghai 200237, China
4. State Key Laboratory of Respiratory Disease, Guangzhou Institute of Oral Disease, Stomatology Hospital, Department of Biomedical Engineering, School of Basic Medical Sciences, Guangzhou Medical University, Guangzhou 511436, China
5. High Magnetic Field Laboratory, Key Laboratory of High Magnetic Field and Ion Beam Physical Biology, Hefei Institutes of Physical Science, Chinese Academy of Sciences, Hefei, China
6. Biomedical Nanocenter, School of Life Science, Inner Mongolia Agricultural University, 29 East Erdos Street, Hohhot, 010011, China

*Corresponding author. Email: fanghaiping@sinap.ac.cn; fengzhang1978@hotmail.com

Abstract:

**There is a considerable amount of work that shows the biomagnetism of organic components without ferromagnetic components at the molecular level, but it is of great challenge to cover the giant gap of biomagnetism between their experimental and theoretical results. Here, we show that the diamagnetism of an aromatic peptide, the AYFFF, is greatly enhanced for about 11 times by self-assembling, reaching two orders of magnitude higher than the mass susceptibility of pure water. Moreover, the AYFFF self-assemblies further mixed with $ZnCl_2$ solution of sufficiently high concentrations display superparamagnetism, with the mass susceptibility reaching more than two orders of magnitude higher than the absolute value of pure water, which may approach the mass susceptibility of ferromagnetism. The aromatic rings in the peptide molecules and the cations are the keys to such a strong diamagnetism and superparamagnetism of aromatic peptides.**

Biomagnetism has attracted tremendous interests in the past few decades.[1] Magnetic fields originating from the heart, brain, skeletal muscles, and isolated nerve and muscle preparations have been detected.[2, 3] High static magnetic fields make cells change their orientation, proliferation, microtubule and mitotic spindle orientation, and their DNA synthesis and cell cycle.[4-6][7] However, most organisms and cells do not contain components that have strong magnetism. Further, it was recently found that self-assembled aromatic peptide nanotubes could align in strong magnetic fields.[8, 9] There have been efforts trying to understand the biomagnetism of organic components without ferromagnetic components. For example, early in 1970s, Worcester and Pauling attributed the diamagnetic property of proteins to aromatic rings and peptide bonds, and estimated the magnitude of diamagnetic susceptibility of these structures.[10, 11] Tendler and Lee argued that, the alignment of aromatic peptide nanotube in magnetic fields was mainly originated from the ordered structure of aromatic rings in the peptide nanotube, given that the aromatic rings have a large diamagnetic anisotropy.[8, 9] However, all their experimental measurements and calculations on the diamagnetic susceptibility are based on the dehydrated samples, partly because the existence of water makes the experimental measurement very difficult. But for the living biological systems, peptides are usually assembled in aqueous solution, and to the best of our knowledge, there is still no report on the magnetism of assembled peptides. On the other hand, it was found that non-dehydrated DNA show a paramagnetic upturn in low temperature, [12] although the mechanism is still unclear.

Here we designed an ingenious experiment to measure the magnetism of assembled aromatic peptides. The diamagnetism of the AYFFF aromatic peptides is greatly enhanced for about 11 times by self-assembling in pure water, with the mass susceptibility reaching two orders of magnitude higher than that of pure water. More unexpectedly, the AYFFF self-assemblies further mixed with $ZnCl_2$ solution of sufficiently high concentrations displays superparamagnetism, which may approach the mass susceptibility of ferromagnetism. It seems that the aromatic rings and the cations are the key to such a strong diamagnetism and superparamagnetism of aromatic peptides. The finding is not only essential for the understanding of the magnetic properties of organic/biological components, but also benefits related applications including peptide assembly in magnetic field, quantitative susceptibility mapping (QSM) and magnetobiology.

The AYFFF aromatic peptide powders were dispersed into pure water (Milli-Q, 18.2 MΩ), namely as the *dispersed state*, where the peptide concentration keeps at 1.0 mg/mL. Then the peptide dispersions were stored still at 20 °C for 16 hours, namely as the *assembled state*. Figure 1d displays the typical self-assembled AYFFF peptide microfibers, which are longer than 10 μm with the width of 300 nm and the height of ~150 nm. For comparison, the AYFFF peptide powders were dissolved in an organic solvent dimethyl sulfoxide (DMSO), in which there is only very few or even negligible amount of assembled peptides. This state is namely as the *dissolved state*.

We used a high-precision electronic balance (XPR205, METTLER TOLEDO, Switzerland) sensitive to 0.01 mg to measure the weight of specimen with or without a fixed magnetic field. 1.0 mL peptide aqueous suspensions of 1.0 mg/mL were filled in a plastic tube, diluted with 1.0 mL pure water, shaken well and placed onto a support on the tray of the electronic balance. The magnetic field was achieved by putting an NdFeB rare-earth permanent magnet at a fixed position on the upper cover of the balance, as shown in Fig. 1a. The readings of the balance with and without the magnet are noted as $w_2$ and $w_1$, respectively. Then the value of ($w_2$-$w_1$) is the magnetic force exerted on the specimen. Here, for simplicity, we used the unit of *mg* (equivalently 9.8 ×10$^{-6}$ N) as the metric for the magnetic force shown in the balance readings. The specimen includes the plastic tube, the tube filled with water/DMSO, the tube filled with water/DMSO and the dissolved peptide, the tube filled with water and the dispersed peptide, and the tube filled with water and the assembled peptide. For each specimen, four samples were prepared and the corresponding forces were measured. Good reproducibility of this magnetic force measurement system has been demonstrated by measuring the magnetic force of the plastic tube, in which the standard deviation from four samples is less than 1%.

We measured the magnetic forces of the plastic tube, plastic tube filled with pure water, plastic tube filled with both peptides at different states and the pure water of the same amount, which were denoted by $w_i^t$, $w_i^{tw}$, and $w_i^{twp}$, $i$=1,2, respectively. Then, the forces

contributed by only water and peptide are

$w_i^w = w_i^{tw} - w_i^t$

and $w_i^p = w_i^{twp} - w_i^{tw}$.

The magnetic forces exerted on the water and peptide are

$w^w = w_2^w - w_1^w$

and $w^p = w_2^p - w_1^p$.

The magnetic force per unit mass of the water and peptide are $w^w/m^w$ and $w^p/m^p$, where $m^w$ and $m^p$ are the mass of water and peptide, respectively.

Figure 1c displays the average magnetic force per unit mass of peptide $<w^p/m^p>$ at different states, together with the average magnetic force per unit mass of water $<w^w/m^w>$ as a reference. It clearly shows that $<w^p/m^p>$ increases in the order of peptides at the dissolved state, at the dispersed state, and at the assembled state. Interestingly, the $<w^p/m^p>$ at the assembled state is about 11 times larger than the value at the dissolved state. This is unexpected since the elements of the peptides at all the above-mentioned different states are the same. For the dispersed state, part of the peptides may be self-assembled so that the value of $<w^p/m^p>$ is between the values at dissolved state, and at assembled state. We can estimate the mass susceptibility of the self-assembled AYFFF peptides, which reaches -1.6×10$^{-6}$ m$^3$/kg, by comparing its value of $<w^p/m^p>$ of ~350 mg/g to the value of $<w^w/m^w>$ of 2.0 mg/g, considering that the mass susceptibility of pure water is well known as -9.1×10$^{-9}$ m$^3$/kg [13].

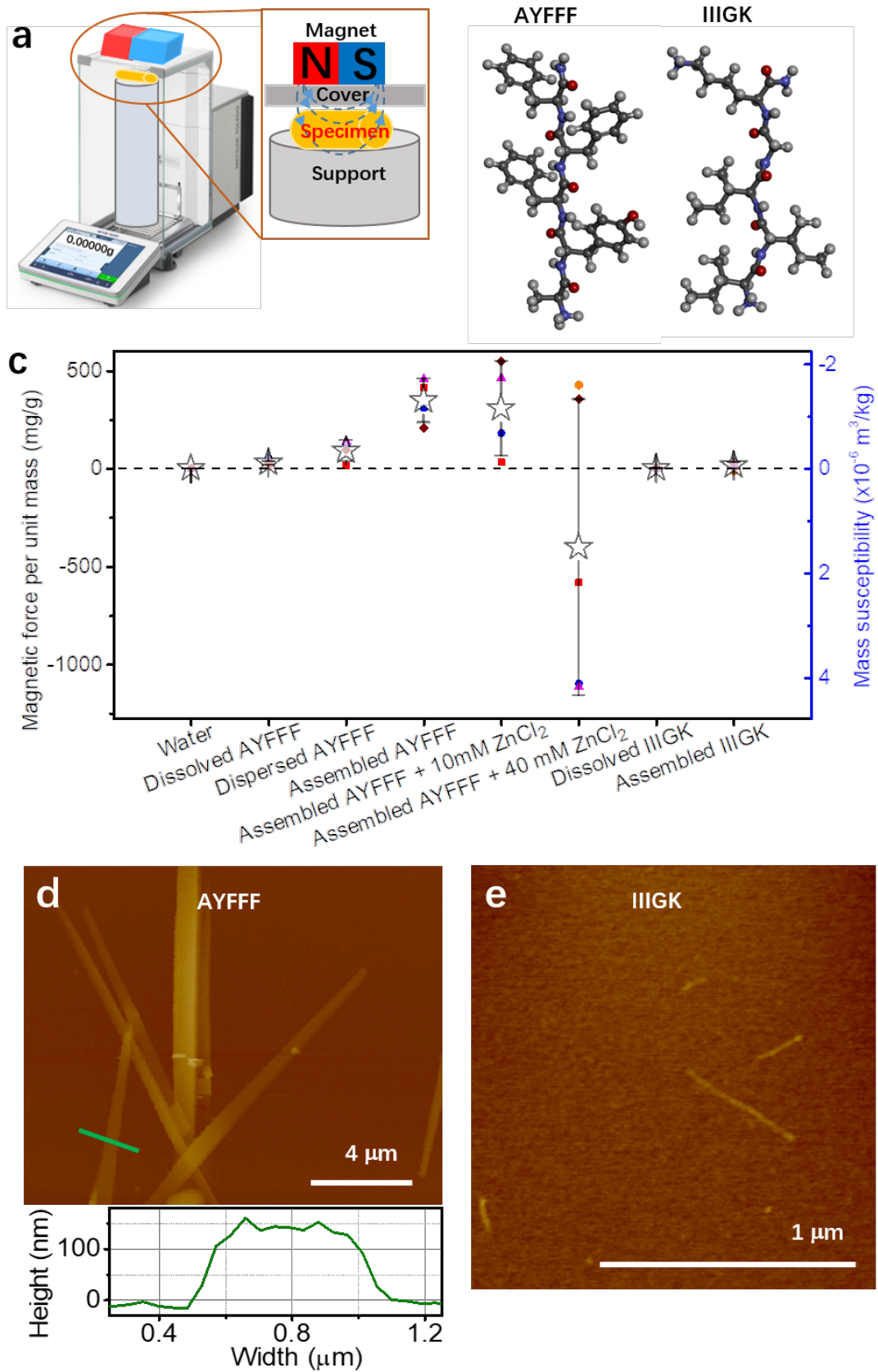

Figure 1. *a) Experimental setup for measuring the magnetism of peptide. The magnet is sitting on the upper glass cover of the electronic balance (inset). The specimens are put on a support placed in the cover of the electronic balance. b) Molecular structures*

*of the AYFFF aromatic peptide and the IIIGK nonaromatic peptide. c) Average magnetic forces per unit mass of the peptides $<w^p/m^p>$ at dissolved, dispersed, assembled states, and assembled AYFFF states further mixed with salt solution of $ZnCl_2$ for different concentrations, respectively, together with water as a reference $<w^w/m^w>$. All the colorful solid scatters display the experimental data, and the grey hollow stars represent their average values with error bars showing their standard deviations. d) AFM images of the self-assembled microfibers of AYFFF aromatic peptides at 20°C in pure water. The height profile corresponding to the green line marked at the position on the AYFFF microfiber. e) AFM images of the self-assembled nanofibers of a control nonaromatic peptide IIIGK. Note: both peptides AYFFF and IIIGK are synthesized with the C-terminal amidation in order to increase their biological activities via generating a closer mimic of native proteins.*

The observation of such strong diamagnetism for the peptides at the assembled state is unexpected. After careful examination of the molecular structure of AYFFF peptide, we presumed that the aromatic rings may be the origin of the strong diamagnetism since the aromatic ring has a large magnetic susceptibility comparing with the other components in the peptides.[10] We then performed measurements with a nonaromatic peptide with the sequence of IIIGK. With the same methods as used for AYFFF peptides, we obtained both the *assembled state* and *dissolved state* of this nonaromatic peptide. The nanofibers with the typical length less than 1 μm and the width of about 10 nm can be seen at the assembled state, as shown in Fig. 1e. The values of $<w^p/m^p>$ of the nonaromatic IIIGK peptide at the assembled state and at the dissolved state were then measured, respectively, both of which are close to $<w^w/m^w>$, the mass susceptibility of pure water. This result confirms that aromatic rings in peptide molecules play the key role in the unexpected strong diamagnetism of aromatic peptides after self-assembling.

Next we consider the salt effect because the biomolecules are usually in saline solution *in vivo*. Taking the salt $ZnCl_2$ as an example, we performed new experiments by only replacing pure water in the previous experiments with salt solutions of different concentration $c(ZnCl_2)$. Explicitly, 1.0 mL AYFFF peptide aqueous suspensions of 1.0 mg/mL were filled in a plastic tube, mixed with 1.0 mL salt solutions, shaken well and placed onto a support on the tray of the electronic balance. Interestingly, as shown in Fig. 1c, the value of $<w^p/m^p>$ decreases with $c(ZnCl_2)$ increasing and becomes negative when $c(ZnCl_2)$ = 40 mM, indicating that the assembled peptides with salts display paramagnetism. Remarkably, the value of $<w^p/m^p>$ of the assembled peptides for $c(ZnCl_2)$ = 40 mM is higher than the absolute value of $<w^p/m^p>$ of the assembled peptides without any salt, indicating the very strong magnetism.

Then we further examined the experimental data at $c(ZnCl_2)$ = 40 mM. The value of $w^p/m^p$ exhibits quite dispersed distribution, ranging from -1112 mg/g to 428mg/g. This is very surprising. We attribute such unexpected behavior to the cation-π interaction between the cation and the aromatic rings in the peptide, where the zinc ions exhibit monovalent behavior as the ferromagnetism observed of the calcium ions on the graphene surface [14]. Clearly, when there is a zinc ion on an aromatic ring, the composition including the ion and the aromatic ring provides a magnetic moment, which may result in a very high paramagnetism and even likely to be ferromagnetism. On the other hand, in such a complex system with the assembled peptide, only a fraction of aromatic rings contains an absorbed zinc ion. If there are only few aromatic rings with an absorbed ion, the whole system is dominated by the aromatic rings without any absorbed ion, and the overall sample would exhibit diamagnetism. On the other hand, when there are enough aromatic rings with absorbed ions, the whole system shows paramagnetism. Clearly, if most of the aromatic rings are absorbed with ions, the assembled peptides will show superparamagnetism and even approach the behavior of ferromagnetism.

In summary, we have experimentally shown that, the diamagnetism of assembled aromatic peptides, which is generally supposed to be comparable to the diamagnetism of water, is unexpectedly strong at the assembling state. Their mass susceptibilities reach two orders of magnitude higher than that of pure water, if the peptides were assembled in pure water. More unexpectedly, the assembled aromatic peptides display paramagnetism with the mass susceptibility more than two orders of magnitude higher than the absolute value of pure water, if the self-assemblies were further mixed with $ZnCl_2$ solution of sufficiently high concentrations, which shows superparamagnetism, or even possibly approach the mass susceptibility of ferromagnetism.

The observation of such strong diamagnetism and superparamagnetism for the aromatic peptides at assembled states is unexpected. It provides opportunities for theoretical investigation and we propose that the self-assembling of aromatic rings in the peptide molecules and the cations with the cation-π interaction, are the keys. Our findings provide a step towards the understanding of the magnetism of biomolecules, as well as the applications such as the magnet induced nano-/microfabrication[15], nuclear magnetic resonance imaging (MRI)[16, 17], and a new pathway for brain-computer interface[18].

## Acknowledgement

This work is supported by the National Natural Science Foundation of China (Nos. U1632135, 11974366, U1932123, 51763019, and U1832125), the Key Research Program of Chinese Academy of Sciences (Grant No. QYZDJ-SSW-SLH053), and the Fundamental Research Funds for the Central Universities.